\documentclass[aps,prb,twocolumn,superscriptaddress,showpacs,ams]{revtex4}
\usepackage{graphicx}

\begin{document}

\title{ Exchange interaction effects in the thermodynamic properties of
quantum dots }
\author{L. G.~G.~V. Dias da Silva}
\email{gregorio@phy.ohiou.edu}
\affiliation{Department of Physics
and Astronomy, Nanoscale and Quantum Phenomena Institute, Ohio
University, Athens, Ohio 45701-2979}
\affiliation{Departamento de
F\'{\i}sica, Universidade Federal de S\~ao Carlos, 13565-905 S\~ao
Carlos SP, Brazil}
\author{Nelson Studart}
\affiliation{Departamento de F\'{\i}sica, Universidade Federal de S\~ao Carlos, 13565-905
S\~ao Carlos SP, Brazil}
\date{\today }

\begin{abstract}
We study electron-electron interaction effects in the thermodynamic
properties of quantum-dot systems. We obtain the direct and exchange
contributions to the specific heat $C_{v}$ in the self-consistent
Hartree-Fock approximation at finite temperatures. An exchange-induced phase
transition is observed and the transition temperature is shown to be
inversely proportional to the size of the system. The exchange contribution
to $C_{v}$ dominates over the direct and kinetic contributions in the
intermediate regime of interaction strength  ($r_{s}\sim 1$). Furthermore,
the electron-electron interaction modifies both the amplitude and the period
of magnetic field induced oscillations in $C_{v}$.
\end{abstract}

\maketitle

\newcommand{\be}   {\begin{equation}}
\newcommand{\ee}   {\end{equation}}
\newcommand{\ba}   {\begin{eqnarray}}
\newcommand{\ea}   {\end{eqnarray}}

\newcommand{\HF}   {\mbox{\scriptsize HF}}
\newcommand{\kin}   {\mbox{\scriptsize kin}}

\section{Introduction}

\label{sec:introduction}

Research on semiconductor quantum dots (QD)s and nanostructures
have drawn considerable effort in recent years.\cite{Mesoreview}
In particular, the study of electron-electron interaction effects
on the ground-state and excited-state of QDs has been a very
active subject. A variety of methods have been used in such
studies, ranging from the exact diagonalization of few electron
systems
\cite{Maksym90,Maksym92,Wagner92,Creffield00,Gregorio02,Sheng_Xu98}
to sophisticated numerical schemes based on the density functional
theory,
quantum Monte Carlo simulations and mean-field approximations.\cite%
{Reimann02} Among the last ones, the self-consistent Hartree-Fock (SCHF)
approximation has been successfully applied to QDs in a number of works \cite%
{Pfannkuche93,Tamura97,Ahn99,Dean01,Reusch03,Vasile_Gudmundsson02,Gregorio04,Szafran04}
which
focused attention on calculations of the pair-correlation function\cite%
{Pfannkuche93} and addition spectra\cite{Tamura97,Ahn99} and on
configurations of the Wigner-like molecule in the strong interacting regime.%
\cite{Reusch03,Szafran04}

A less pursued track is the use of SCHF to study magnetic and
thermodynamic properties of semiconductor QDs.
\cite{Vasile_Gudmundsson02,Dean01,Gregorio04} Electron-electron
interaction was shown to give an important contribution to
thermodynamic properties such as the magnetization
\cite{Sheng_Xu98,Vasile_Gudmundsson02} and the magnetic
susceptibility.\cite{Gregorio02,Gregorio04} Another quantity of
experimental interest is the specific heat $C_{v}$, which has been
studied in a number of works in both non-interacting
\cite{DeGroote92,Geyler97} and interacting \cite{Maksym90,Dean01}
QD systems. In Ref. \onlinecite{Dean01}, Dean and coworkers
reported an interesting interaction-induced phase transition in
parabolic QDs with $N\sim 6$ electrons. This phase transition
manifests itself as sharp drops in the specific heat as the
temperature reaches a critical value. Nevertheless, a systematic
study on how such a transition depends on the interaction coupling
parameter $r_{s}$ of the dot, which measures the relative
electron-electron interaction strength, remains to be performed.

In this paper, we address the role of the exchange interaction in the
thermodynamic properties of non-parabolic QDs. Specifically, we study the
kinetic, direct and exchange contribution to the specific heat in a
finite-temperature Hartree-Fock approach.\cite{Tamura97,Ahn99} In a previous
work using this method, \cite{Gregorio04} we have shown that the exchange
interaction contribution is the dominant term in magnetic properties such as
the zero-field susceptibility in the intermediate regime of interaction
strength ($r_{s}\sim 1$). We find in this investigation that the exchange
effects also play a dominant role on the specific heat properties. In
particular, the exchange electron correlations dominate the
finite-temperature phase transition and is the leading contribution to $C_{v}
$ for $r_{s}\sim 1$. We also find that the transition temperature scales
with the inverse of the dot size. As a consequence, this phase transition
could in principle be experimentally observed for dots tens of nanometers
across at an attainable temperature range.

The paper is organized as follows. In Section \ref{sec:refsys} we describe
the system to be studied and its main features. The results for the specific
heat and the discussion of main results are given in Section \ref%
{sec:results} as well as our closing remarks.

\section{The Model}

\label{sec:refsys}

We consider the problem of $N$ interacting electrons confined in a 2D square
quantum dot of size $L$ and subjected to an external magnetic field $\mathbf{%
B}$ perpendicular to the electron system. To account for screening effects,
the electron-electron interaction is modeled by an Yukawa-type potential%
\cite{Gregorio02} and the model Hamiltonian reads as%
\begin{equation}
H=\sum_{n=1}^{N}h(\mathbf{r}_{n})+\sum_{n<{n^{\prime }}}^{N}\frac{e^{2}}{%
\epsilon _{r}}\frac{e^{-\kappa |\mathbf{r}_{n}-\mathbf{r}_{n^{\prime }}|}}{|%
\mathbf{r}_{n}-\mathbf{r}_{n^{\prime }}|}\,,
\end{equation}%
where $\mathbf{r}_{n}$ indicates the position of the $n$th
electron. We consider low $g$-factor QDs, so that the Zeeman term
can be safely disregarded. Above, $\kappa $ gives the effective
interaction range and $\epsilon _{r}$ is the background dielectric
constant. For $\kappa =0$, there are no screening effects and the
\textquotedblleft bare\textquotedblright\ Coulomb interaction is
recovered.

The single-particle Hamiltonian $h(\mathbf{r})$ is given by
\begin{equation}
h(\mathbf{r})=\frac{1}{2m^{\ast }}\left[ \mathbf{p}+\frac{e}{c}\mathbf{A}(%
\mathbf{r})\right] ^{2}+u(\mathbf{r})\,,  \label{eq:1pHamiltonian}
\end{equation}%
where $m^{\ast }$ is the electron effective mass and $u(\mathbf{r})$ is the
hard-wall confining potential. The vector potential $\mathbf{A}$ is chosen
in the symmetric gauge, namely, $\mathbf{A}=(-By/2,Bx/2,0)$. Hereafter, the
magnetic field is expressed in units of $\Phi /\Phi _{0}$, where $\Phi =B%
\mathcal{A}$ is the magnetic flux through the system area $\mathcal{A}$ and $%
\Phi _{0}=hc/e$ is the quantum flux unit.

A key parameter in our analysis is $L/a_{B}^{\ast }$, the QD length $L$ in
units of the effective Bohr radius $a_{B}^{\ast }=\hbar ^{2}\epsilon _{r}/{%
m^{\ast }e^{2}}$ which gives the relative strength of the e-e
interaction as compared to the kinetic energy of the
system.\cite{Ahn99} For a square dot of side $L$, the potential
energy scales with $L^{-1}$ while the kinetic energy scales with
$L^{-2}$. Therefore, as $L$ is increased, the potential energy
becomes increasingly more important.

The standard dimensionless parameter that quantifies the ratio between the
potential and kinetic energies of the system is the so-called Brueckner
parameter $r_{s}$, which in 2D reads as $r_{s}^{2}=\mathcal{A}/(N\pi \lbrack
a_{B}^{\ast }]^{2})$. Therefore, $L/a_{B}^{\ast }$ and $r_{s}$ are related by $%
r_{s}=(L/a_{B}^{\ast })/\sqrt{\pi N}$. Furthermore, by choosing a square
hard-wall confinement, one can easily tune $r_{s}$ by changing the dot's
lateral size $L$.

The many-body ground-state energy is obtained in the finite-temperature SCHF
approximation. The SCHF equations read as\cite{Tamura97,Dean01}
\begin{eqnarray}
h({\bf r}) \phi_i({\bf r}) + \sum_j \left[ n_j \!
   \int d{\bf r}^\prime \phi_j^*({\bf r}^\prime)v( {\bf r},{\bf r}^\prime)
   \phi_j({\bf r}^\prime)\right]\phi_i({\bf r}) \nonumber \\
 - \sum_j \left[ n_j\! \int\!\! d{\bf r}^\prime \phi_j^*({\bf r}^\prime)
   v({\bf r},{\bf r}^\prime)\phi_j({\bf r})\phi_i({\bf r}^\prime) \right]
 = \varepsilon_i^{\HF} \phi_i({\bf r}) ,
\label{eq:HFeq}
\end{eqnarray}
where the sums run over all HF orbitals. Here $n_{i}=\{\exp
[(\varepsilon _{i}^{\mathrm{HF}}-\mu )/k_{B}T]+1\}^{-1}$ is the
Fermi occupation number of the $i$th HF orbital with corresponding
wave function $\phi _{i}(\mathbf{r})$ and energy $\varepsilon
_{i}^{\HF}$. As in the standard
procedure, the chemical potential $\mu $ is determined by requiring that $%
N=\sum_{i}n_{i}$. We truncate the number of orbitals and take only the $%
M\geq 2N$ lowest energy states into account.

The SCHF ground-state energy is given by
\begin{eqnarray}
E_{g}^{\HF} &\equiv &T^{\HF}+V_{d}^{%
\HF}-V_{x}^{\HF}  \nonumber \\
&=&\sum_{i}n_{i}\langle \phi _{i}|h|\phi _{i}\rangle +\frac{1}{2}%
\sum_{i,j}n_{i}n_{j}\Big(\langle \phi _{i}\phi _{j}|v|\phi _{i}\phi
_{j}\rangle   \nonumber \\
&&-\langle \phi _{i}\phi _{j}|v|\phi _{j}\phi _{i}\rangle \Big),
\label{eq:EHFdireto}
\end{eqnarray}%
where the $|\phi _{i}\rangle $ are the HF orbitals, self-consistent
solutions of Eq.\ (\ref{eq:HFeq}) and $T^{\HF}$, $V_{d}^{%
\HF}$ and $V_{x}^{\HF}$ are the kinetic, direct and exchange
contributions to the ground-state energy respectively. We are
interested in the intermediate interaction strength regime
($r_{s}\sim 1$). In this regime, the direct interaction term
$V_{d}^{\HF}$ is the leading contribution to the $E_{g}^{\HF}$,
followed by $T^{\HF}$ and $V_{x}^{\HF}$ respectively.

The details on the matrix elements calculations and the Hartree-Fock method
can be found on Refs. \onlinecite{Gregorio02} and \onlinecite{Gregorio04}
respectively.

We calculate the specific heat $C_{v}$ of the system in this SCHF
approximation, namely:

\begin{equation}
C_{v}=\left( \frac{\partial U}{\partial T}\right) _{V}
\end{equation}%
where $U$ is the internal energy and $T$ is the system temperature. The
first-order exchange and Hartree contributions to $C_{v}$ can be accounted
for by approximating $U\approx E_{g}^{\HF}$ so that for $%
E_{g}^{\HF},$ given by Eq. \ref{eq:EHFdireto}, there are kinetic
$(C_{v}^{\kin})$, direct $(C_{v}^{d})$ and exchange $(C_{v}^{x})$
contributions to $C_{v}$.

\section{Results and Conclusions}

\label{sec:results}

We analyze the behavior of the specific heat $C_{v}$ as a function of
relevant parameters of the system, i.e. the temperature $T$, the interaction
strength $L/a_{B}^{\ast }$, the magnetic field $\Phi /\Phi _{0}$, the
potential range $\kappa ^{-1}$ and the number of electrons $N$. The results
given in this section are for the Coulomb case $(\kappa =0)$ with $N=10$
electrons in the dot. We should mention that QDs with up to $N=20$ electrons
and with the screened interaction $(\kappa \neq 0)$ were also considered and
the same overall qualitative features were observed. In the remaining of
this section, energy and temperature are given in units of the typical
scales for the system, namely $E_{L}=\hbar ^{2}/(m^{\ast }L^{2})$ and $%
T_{L}=E_{L}/k_{B}$ respectively.

\begin{figure}[h]
\includegraphics*[width=1\linewidth]{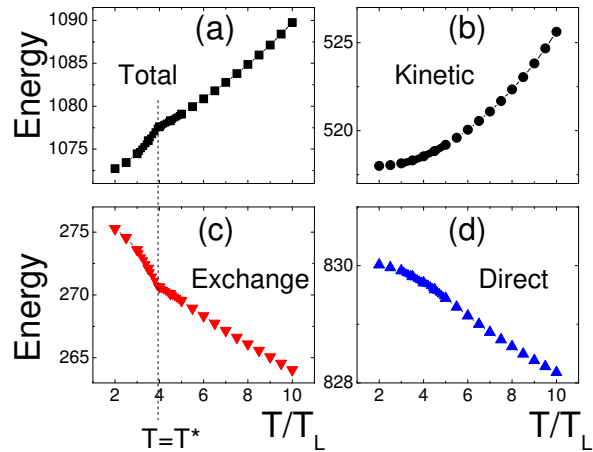}
\caption{ (color online) (a): Ground-state energy $E_{\HF}$ as
function of temperature for $N=10$ electrons and $r_s=0.89$. The
kinetic (b), exchange (c) and direct (d) contributions to
$E_{\HF}$ are also plotted. }
\label{fig:EHF_T}
\end{figure}

The ground-state energy $E_{g}^{\HF}$ increases with temperature
as shown in Fig. \ref{fig:EHF_T}a for $N=10$ and $r_{s}=0.89$.
Nevertheless, this increase is not smooth and a sudden change in
slope is observed at a certain temperature $T^{\ast }$. By
analyzing the kinetic
energy $T^{\HF}$, the direct term $V_{d}^{%
\HF}$ and the exchange contribution $V_{x}^{%
\HF}$  given by Eq. (\ref{eq:EHFdireto}), we see that both
$V_{d}^{\HF}$ and $V_{x}^{\HF}$ decrease with $T$, as expected.
Furthermore, we observe clearly that the change in
slope is a feature due to the exchange interaction, since neither $T^{%
\HF}$ or $V_{d}^{\HF}$ display cusps at $%
T=T^{\ast }$ (Figs. \ref{fig:EHF_T}b-d).

\begin{figure}[tbp]
\includegraphics*[width=1\linewidth]{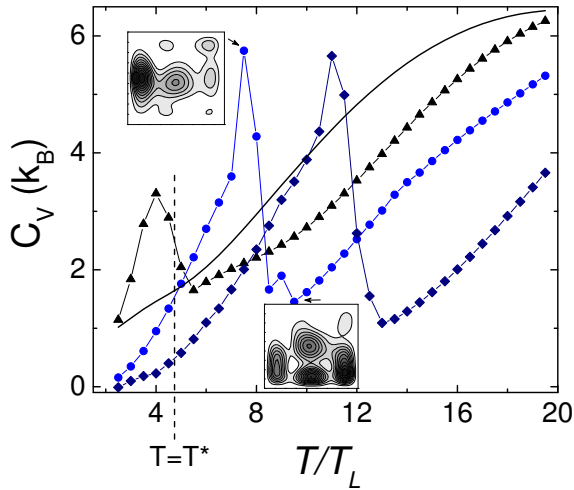}
\caption{ (color online) Specific heat as function of temperature
for the noninteracting (solid line) and interacting cases. In the
latter, the interaction strength values are $r_s=1.07$ (up
triangles), $2.14$ (circles) and $3.30$ (diamonds). The insets show
typical charge distributions in the dot before and after the
transition. } \label{fig:Cv_TB}
\end{figure}

The abrupt change in slope in the energy causes a discontinuity in
the specific heat $C_{v}(T)$ at $T=T^{\ast }$, as seen in Fig.
\ref{fig:Cv_TB}. A sharp drop develops for a wide range of values of
the interaction strength parameter (for simplicity, the
noninteracting case is referred to as \textquotedblleft
$L/a_{B}^{\ast }=0$\textquotedblright\ or \textquotedblleft $r_{s}=0$%
\textquotedblright . The noninteracting curve does not display any sharp
drops).

Discontinuities in the specific heat are usually regarded as signatures of
phase transitions.\cite{Stanley} In fact, such transitions are accompanied
by a charge reordering in the ground state distribution, as shown in the
insets of Fig. \ref{fig:Cv_TB}. The lack of the $\mathcal{C}_{4}$ rotational
symmetry in the charge distributions is a consequence of the nonlinear
coupling of the original orbitals in the Eq. (\ref{eq:HFeq}) and is a
peculiarity of the Hartree-Fock approximation. Nonetheless, a clear charge
rearrangement is verified as the system undergoes the phase transition. Most
strikingly is the fact that this is an exchange-induced phase transition and
it is a direct consequence of exchange effects between the electrons in the
dot. Such transitions were reported in previous studies \cite{Dean01} as
related to phase transitions in the ground-state charge distribution inside
the dot.

\begin{figure}[tbp]
\includegraphics*[width=1\linewidth]{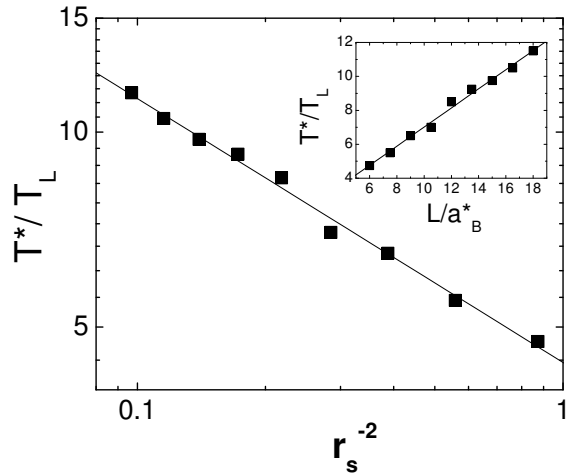}
\caption{ Reduced transition temperature $T^{*}/T_L$ as function of
$r_s^{-2}$ (see text for details). Inset: $T^{*}/T_L$ increases
linearly with the interaction strength $L/a^{*}_B$. }
\label{fig:Tstar_La}
\end{figure}

The reduced transition temperature varies both as a function of
the number of electrons and the relative interaction strength. For
$N=20$, up to three transitions are observed in the temperature
range $2<T/T_{L}<20$ (not shown). In Fig. \ref{fig:Tstar_La}, the
dependence of the critical temperature $T^{\ast }/T_{L}$ with\
$r_{s}^{-2}$ (which is proportional to the density $N/L^{2}$) is
shown. We also depict $T^{\ast }/T_{L}$ as a function of
$L/a_{B}^{\ast }$ in the inset of Fig. \ref{fig:Tstar_La} and a
roughly linear dependence is observed. These results imply that
the transition temperature $T^{\ast }$ scales with $L^{-1}$ since
$T_{L}\propto L^{-2}$. The data are very well described by a
linear fit ($T^{\ast }\propto L^{-1}$) as shown in the figure.

Such a scaling behavior allows us to estimate the values of the
critical temperature $T^{\ast }$ for typical dot sizes. For GaAs QDs
with $L\sim 50$ nm one obtains $T^{\ast }\sim 11$ K and it decreases
with $1/L$ for larger dots. The transition temperature further
decreases when screening effects
are taken into account. For the potential range of $\kappa ^{-1}=L/10$ and $%
L\sim 50$ nm, we obtain $T^{\ast }\sim 2$ K.

\begin{figure}[tbp]
\includegraphics*[width=1\linewidth]{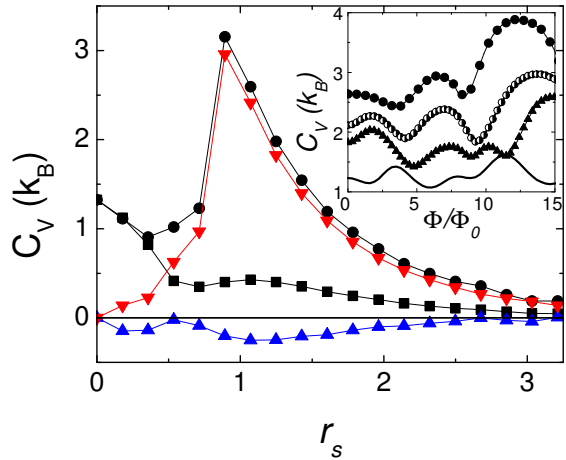}
\caption{ (color online) Specific heat as function of $r_s$ (circles). Also
plotted are the kinetic (squares), direct (up triangles) and exchange (down
triangles) contributions. Inset: $C_v$ as function of magnetic flux $%
\Phi/\Phi_0$ for the noninteracting (solid line) and interacting
cases with $r_s=0.53$ (triangles), $0.89$ (half-filled circles)
and $1.78$ (filled circles). Curves are off-set for clarity.}
\label{fig:Cv_La}
\end{figure}

We have also investigated the specific heat dependence with both the
interaction strength parameter and the magnetic field for a fixed
temperature. The kinetic, direct, and exchange contributions to $C_{v}$ at $%
T=3T_{L}$ are shown on Fig. \ref{fig:Cv_La} as a function of
$r_{s}$. As the relative interaction strength parameter increases,
the exchange contribution $C_{v}^{x}$ rises fast and becomes the
leading contribution to $C_{v}$ for $r_{s}\sim 1$. This is a
surprising result since $V_{x}^{\HF}$ is smaller than
$V_{d}^{\HF}$ and $T^{\HF}$ by a factor of $2-3$ (see Fig.
\ref{fig:EHF_T}). However, the effect of temperature in the
variation of $V_{x}^{\HF}$ is stronger
and $C_{v}^{x}>C_{v}^{\kin},$ $C_{v}^{d}$ for $r_{s}\sim 1$%
. The direct term, on the other hand, gives a smaller negative
contribution which cancels out the positive $C_{v}^{\kin}$ (which
is dominant for $r_{s}\ll 1$). A peak in $C_{v}$ appears at the
value of $r_{s}$
for which $T^{\ast }=3T_{L}$ and approaches zero for higher values of $r_{s}$%
, since temperatures changes do not sensibly affect the ground-state energy
in the strongly interacting regime.

The specific heat oscillates as a function of the magnetic field with
increasing amplitude, as seen in the inset of Fig. \ref{fig:Cv_La}. The
interaction influences both the period and the amplitude of $C_{v}(B)$. The
noninteracting curve displays an oscillatory pattern with both high and low
harmonics. For higher values of the interaction strength parameter, the
higher harmonics are suppressed and an oscillation period is defined more
clearly . For even higher values of $r_{s}$, the oscillation period
decreases. A similar behavior was seen in the magnetization and magnetic
susceptibility of QD systems\cite{Gregorio04} and it is related to an
effective increase in the chemical potential as $r_{s}$ increases.

In summary, we have investigated interaction effects in the thermodynamic
properties of QDs. The exchange interaction plays a relevant role on the
specific heat features and is the leading contribution for dots in the $%
r_{s}\sim 1$ range. The exchange induced finite-temperature phase
transition, studied in previous works,\cite{Dean01} has been shown
to depend on the interaction strength parameter, or equivalently,
on the system size $L $. The transition temperature $T^{\ast }$
decreases as $L$ increases and we estimate that the transition
regime could be experimentally accessible for sufficiently small
dots. Furthermore, the specific heat oscillates with the magnetic
field and both period and amplitude of such oscillations strongly
depend on interaction effects.

Experiments to verify our findings using single quantum-dots are
likely too demanding. The specific heat has been measured in
multi-layer 2D electron gases in the Landau regime with heat-pulse
\cite{Gornik85} and steady-state ac-temperature calorimetry
\cite{Wang88} techniques, with resolutions in $C_v$ still much
lower than the required to test our results. One possible way to
overcome such difficulty is to perform experiments in ensembles of
nearly identical dots in a multi-layer configuration so that the
contribution from single dots is amplified. This is, nonetheless,
an experimentally challenging task which would bring a new
understanding to the many-body effects in the thermodynamics of
such small-scale devices.

\acknowledgments

We thank Profs. Caio Lewenkopf and Sergio Ulloa for helpful comments and
suggestions. This work was partly supported by FAPESP (grant 01/14276-0),
CNPq and NSF.

\end{document}